\documentclass{emulateapj}

\received{2013 November}
\slugcomment{To appear in {\it the Astrophysical Journal, Letter}}

\shortauthors{KOBAYASHI et al.}
\shorttitle{The origin of low [$\alpha$/Fe] ratios in extremely metal-poor stars}

\def\gtsim {>\kern-1.2em\lower1.1ex\hbox{$\sim$}~}   
\def\ltsim {<\kern-1.2em\lower1.1ex\hbox{$\sim$}~}   

\begin{document}

\title{The origin of low [$\alpha$/Fe] ratios in extremely metal-poor stars}
\author{Chiaki KOBAYASHI$^{1,2}$, Miho N. ISHIGAKI$^{2}$, Nozomu TOMINAGA$^{2,3}$,
and Ken'ichi NOMOTO$^2$}
\affil{$^1$ School of Physics, Astronomy and Mathematics,
Centre for Astrophysics Research, University of Hertfordshire,
College Lane, Hatfield  AL10 9AB, UK; c.kobayashi@herts.ac.uk}
\affil{$^2$ Kavli Institute for the Physics and Mathematics of the Universe (WPI), The University of Tokyo, Kashiwa, Chiba 277-8583, Japan}
\affil{$^3$ Department of Physics, Faculty of Science and Engineering, Konan University, 8-9-1 Okamoto, Kobe, Hyogo 658-8501, Japan}

\begin{abstract}
We show that the low ratios of $\alpha$ elements (Mg, Si, and Ca) to Fe recently found for a small fraction of extremely metal-poor stars can be naturally explained with the nucleosynthesis yields of core-collapse supernovae, i.e., $13-25M_\odot$ supernovae, or hypernovae.
For the case without carbon enhancement, the ejected iron mass is normal, consistent with observed light curves and spectra of nearby supernovae.
On the other hand, the carbon enhancement requires much smaller iron production, and the low [$\alpha$/Fe] of carbon enhanced metal-poor stars can also be reproduced with $13-25M_\odot$ faint supernovae or faint hypernovae.
Iron-peak element abundances, in particular Zn abundances, are important to put further constraints on the enrichment sources from galactic archaeology surveys.
\end{abstract}


\section{Introduction}

The observed elemental abundances of metal-poor stars can be used to constrain the physics of supernovae \citep[see][for a review]{nom13}.
Although the nature of the first stars is not well understood, if they explode as supernovae the first chemical enrichment is imprinted in the elemental abundances of the second generation of stars.
During the early stages of galaxy formation, the interstellar medium (ISM) is highly inhomogeneous, and it is likely that these elemental abundance patterns are determined only by a few supernovae \citep{aud95}.
In the inhomogeneous enrichment, metallicity is not a time indicator
anymore but merely reflects the metallicity of the cloud in which the
second generation of stars formed.
This metallicity is often estimated with an analytic formula \citep{tom07}, but detailed hydrodynamical simulations with radiative cooing are necessary to predict the metallicity distribution function of the second generation of stars.
Alternatively, we assume that metal poor stars with [Fe/H] $\ltsim -3$ are enriched by a single supernova, and study the properties of the supernova by comparing observed elemental abundances and nucleosynthesis yields (abundance profiling/fitting).

Thanks to large scale surveys and follow-up high resolution
spectroscopy, intensive observations of metal-poor stars have
revealed the existence of extremely-, ultra-, and hyper- metal-poor (EMP, UMP, and HMP) stars 
with [Fe/H] $=(-4,-3),(-5,-4),(-6,-5)$, respectively \citep{bee05}.
The elemental abundance patterns have a distinct signature:
Including two HMP stars, $10-25$\% of stars with [Fe/H] $\ltsim -2$ show carbon enhancement relative to iron ([C/Fe] $\gtsim1$, \citealt{aok10}).
Such carbon-enhanced metal-poor (CEMP) stars often show enhancement of $\alpha$ elements (O, Mg, Si, S, and Ca), although one star at [Fe/H]$=-4.99$ does not show enhancement of C, N, and Mg \citep{caf11}.
There are various scenarios to explain the carbon enhancement, including rotating massive stars \citep{mey06}, asymptotic giant branch stars in binary systems \citep{sud04,lugaro08}, and black-hole-forming core-collapse supernovae (faint supernovae; \citealt{ume02,iwa05,tom07}).
\citet{caf13} recently presented different types of EMP stars that show lower [$\alpha$/Fe] ratios with/without carbon enhancement than other EMP stars.
The variation of [$\alpha$/Fe] ratios of EMP stars has been known. 
As the quality of data improved, \citet{cay04} concluded that without CEMP, the scatter of elemental abundances is so small that the ISM is well mixed at the early stages of galaxy formation.
However, a significant scatter is seen in other observational data \citep[e.g.,][]{hon04,yon13,coh13} and a small fraction of stars show lower [$\alpha$/Fe] ratios than $\sim$ 0.2.

An intrinsic variation of [$\alpha$/Fe] ratios can be caused from the following enrichment sources:
(i) The most popular source is Type Ia Supernovae (SNe Ia), which produce more Fe than $\alpha$ elements.
Depending on the fractional contribution of core-collapse supernovae from previous populations, [$\alpha$/Fe] can vary between $\sim 0.5$ to $\sim -0.6$.
However, there is a time-delay of the enrichment in the case of SNe Ia, which depends on the progenitor systems \citep{kob09}, and is $\sim 34$ Myr at the shortest for $8 M_\odot$ primary stars.
Hence, it is unlikely that many EMP stars are affected by SNe Ia.
(ii)
$\sim 10-20 M_\odot$ supernovae have a smaller mantle mass that contains $\alpha$ elements than more massive stars, and thus give lower [$\alpha$/Fe] ratios than the initial mass function (IMF) weighted values of core-collapse supernova yields, i.e., the plateau values of [$\alpha$/Fe]-[Fe/H] relations \citep{kob06,kob11agb}.
These supernovae will leave a neutron star behind, and should be very common for the standard IMF weighted for the low-mass end \citep[e.g.,][]{kro08}.
(iii) 
Hypernovae ($E_{51} \equiv E/10^{51} {\rm erg}\gtsim10$ for $\gtsim25M_\odot$) are observationally known to produce more iron than normal supernovae ($E_{51}\sim1$ for $\gtsim10M_\odot$) \citep[e.g.,][]{nom03,sma09}.
Therefore, hypernovae can give lower [$\alpha$/Fe] ratios than supernovae at a given progenitor mass.
The hypernova rate is not very high at present, but can be high for low-metallicity stars because of small angular momentum loss.
(iv) 
Faint supernovae are proposed to explain the elemental abundance patterns of CEMP stars from carbon to zinc.
The central parts of supernova ejecta that contain most of iron fall back onto the black hole, while the stellar envelopes that contain carbon are ejected as in normal supernovae.
Therefore, the [C/Fe] ratio of faint supernovae is as large as that of CEMP stars.
Among $\alpha$ elements, O and Mg are synthesized during hydrostatic burning and located in the outskirts of ejecta.
Therefore, faint supernovae often have high [(O, Mg)/Fe] ratios, depending on mixing-fallback processes.
The faint supernovae scenario is also the best explanation of the observed carbon-enhanced damped Lyman $\alpha$ (DLA) system \citep{kob11dla}.
(v)
Primordial stars with initial masses of $\sim 140-270 M_\odot$ enter into the electron-positron pair-instability region during the central oxygen-burning stages, where most of O and Mg are transformed into Si, S, and Fe.
Pair-instability supernovae produce a much larger amount of iron and higher [(Si,S)/(O,Mg)] ratios than core-collapse supernovae.
Such abundance patterns have been found neither in EMP stars \citep[e.g.,][]{cay04} nor in DLA systems \citep{kob11dla}.

In this Letter, we explore our supernova and hypernova models with/without mixing-fallback over a wide range of progenitor mass.
We then perform abundance fitting to observed low-$\alpha$ stars and discuss the enrichment sources (\S2).
In \S3 we give more general discussion on supernovae and chemical enrichment, and  summarize our main conclusions in \S4.

\begin{table*}[tbh]
\begin{center}
\begin{tabular}{lccccccccccc}
\hline
Name  & [Fe/H] & [Mg/Fe] & [Ca/Fe] & $M$ & $E_{51}$&  $M_{\rm cut}$ & $M_{\rm mix}$  & $\log f$& $M_{\rm rem}$ &  $M(^{56}{\rm Ni})$  & $\chi^{2}/N$\\
      & (dex) & (dex) & (dex) & ($M_{\odot}$)&($10^{51}$ erg) &  ($M_{\odot}$) & ($M_{\odot}$)  &    & ($M_{\odot}$) &  ($M_{\odot}$)    &\\
\hline
J144256$-$001542 & $-4.09\pm0.21$ & $0.27$ & $0.29$ & 15 & 1  & 1.5  & -   &  $0.0$ &  1.5  &  0.06   &0.12\\
                      &                &        &        & 15 & 1  & 1.4  & 1.5 & $-1.2$ &  1.5  &  0.06  &0.12\\
                      &                &        &        & 25 & 1  & 1.7  & 2.9 & $-0.4$ &  2.4  &  0.10  &$<$0.01\\
                      &                &        &        & 40 & 30 & 2.2  & 5.9 & $-0.8$ &  5.3  &  0.33  &$<$0.01\\
J153346$+$155701 & $-3.34\pm0.26$ & $0.06$ & $0.08$ & 15 & 1  & 1.4  & -   &  $0.0$ &  1.4  &  0.14   &0.38\\
                      &                &        &        & 15 & 1  & 1.4  & 1.5 & $-0.3$ &  1.4  &  0.09     &0.09\\
                      &                &        &        & 25 & 1  & 1.7  & 3.6 & $-0.3$ &  2.6  &  0.13     &0.19\\
                      &                &        &        & 25 & 10 & 1.8  & 3.5 & $-0.4$ &  2.8  &  0.24     &$<$0.01\\
J161956$+$170539 & $-3.57\pm0.25$ & $0.04$ &$-0.35$ & 15 & 1 &  1.4 & 3.2 & $-4.0$ & 3.2 & 1.4$\times 10^{-5}$   &0.30\\ 
                      &                &        &        & 25 & 10 &  1.7 & 6.5 & $-4.0$ & 6.5 & 6.7$\times 10^{-5}$  &0.71\\ 
\hline
HE0305-5442 & $-3.30\pm0.20$ & $0.22$ & $-0.04$ & 15 & 1  & 1.4  & 1.6   &  $-0.5$ &  1.5  &  0.04   &1.37\\
                      &                &        &        & 25 & 10 & 1.7  & 6.2& $-1.6$ &  6.1 &  0.02     &1.14\\
HE1416-1032 & $-3.20\pm0.16$ & $0.18$ & $0.03$ & 15 & 1  & 1.4  & 4.3   &  $-2.4$ &  4.3  &  5.4$\times 10^{-4}$   &3.10\\
                      &                &        &        & 25 & 10 & 1.9  & 9.1 & $-3.6$ &  9.1 &  1.3$\times 10^{-4}$     &3.67\\
HE2356-0410 & $-3.06\pm0.20$ & $0.11$ &$0.16$ & 15 & 1 &  1.4 & 3.1 & $-3.7$ & 3.1 & 2.7$\times 10^{-5}$   &8.64\\ 
                      &                &        &        & 25 & 10 &  1.9 & 7.1 & $-4.7$ & 7.1 & 1.0$\times 10^{-5}$  &6.01\\ 
\hline
\end{tabular}
\caption{\rm Observed abundances, the parameters of nucleosynthesis models (progenitor mass $M$, explosion energy $E_{51}$, inner boundary $M_{\rm cut}$, outer boundary $M_{\rm mix}$, and ejection fraction $f$ of the mixing region), the outputs (remnant mass $M_{\rm rem}$, and ejected iron mass M$(^{56}{\rm Ni})$), and the $\chi^{2}$ of the abundance fitting.}
\end{center}
\end{table*}

\section{Abundance Fitting}

We adopt progenitor star models from \citet{ume05} for $13, 15, 25$ and $40M_\odot$ with the same modification of the number of electrons per nucleon, $Y_{\rm e}$, as in \citet{kob06}.
During the supernova explosion, the elements synthesized in different stellar layers should mix to some extent, and some fraction of this mixed material falls back onto the remnant.
We calculate nucleosynthesis yields by using mixing-fallback models 
\citep{ume02,iwa05,tom07} for supernovae and hypernovae.
The mass-energy relation is taken from the results of light curve fitting \citep{nom03}; $E_{51}=1$ for $13-40M_\odot$ supernovae, and $E_{51}=10$ and $30$ for $25$ and $40M_\odot$ hypernovae, respectively.
We then perform the parameter search for three dimensional spaces of the inner boundary of mixing $M_{\rm cut}$ (i.e. mass cut), the outer boundary of mixing $M_{\rm mix}$, and the ejection fraction of the mixing region $f$.
The minimum, maximum, steplength for each parameter are: $M_{\rm cut}=$ [iron core mass, $3M_\odot$, $0.1M_\odot$], $M_{\rm mix}=$ [$M_{\rm cut}$, MIN(CO core mass, $16M_\odot$), $0.1M_\odot$], and $\log f=[-5, 0, 0.1]$.
Note that $f=1$ corresponds to the case without mixing-fallback.
We search acceptable models using the chi-squared test statistic, 
and the parameters of favoured models that have least $\chi^{2}$ for given $M$ and $E$ are listed in Table 1.
Acceptable models have parameters around the peaks and give similar $\chi^{2}$ values.

For the three stars in \citet{caf13}, 
the observational constraints are [Mg/Fe], [Si/Fe], and [Ca I/Fe], and detection or non-detection of [C/Fe].
Provided signal-to-noise ratio, temperature, and gravity of the two stars with no carbon detection, we estimate that [C/Fe] ratios should be lower than $\sim 1.5$ from spectral synthesis, and apply this constraint.
It is the iron peak elements that give a much stronger constraint on model parameters, but there are no observational estimates available for these stars.
Instead, we put a constraint on the Ni abundance as [Ni/Fe] $<0.5$ because [Ni/Fe] is almost 0 for a wide range of metallicity (see Fig. 24 of \citealt{kob06}).
Figure 1 shows the comparison of the elemental abundance patterns between observations and favored models.

{\it SDSSJ144256+001542} --- The observed [$\alpha$/Fe] ratios ($\sim 0.3$) 
are consistent with LTE abundances of typical halo stars \citep{coh13,yon13}, but are slightly lower than the IMF-weighted yields ([Mg/Fe] $\sim 0.5$\footnote{This is consistent with NLTE abundances \citep{and10}.} and [Ca/Fe] $\sim 0.3$).
Therefore, the observed [$\alpha$/Fe] ratios can be reproduced with the yields of (1) $13-15M_\odot$ supernovae, (2) $13-40M_\odot$ supernovae with mixing-fallback, or (3) hypernovae. 
Among normal supernovae (with ejected iron mass $M({\rm Fe}) \sim 0.07M_\odot$), if the progenitor mass is as low as $13-15M_\odot$, the observed ratios can be reproduced even without mixing-fallback (solid line, see \S 1).
However, $25-40M_\odot$ models without mixing-fallback give $M({\rm Fe})$ larger than observed supernovae.
The remnant mass is around $M_{\rm rem} \sim 2M_\odot$, which is in the range of neutron stars.
With mixing-fallback, the observed ratios can be reproduced with many parameter sets.
However, to be consistent with observed supernovae ($M({\rm Fe})=0.05-0.15M_\odot$), $40M_\odot$ models require extended mixing and large fallback ($M_{\rm mix}>8 M_\odot$ and $f<0.3$), in order for a large fraction of the CO core to fall back, which is not very realistic (see \S 3).
For hypernovae (dotted line), as we allow $M({\rm Fe})>0.1M_\odot$, [$\alpha$/Fe] can be reproduced with $25-40M_\odot$.
The remnant mass is $M_{\rm rem} \gtsim 4M_\odot$, which is in the range of black holes.

{\it SDSSJ153346+155701} --- Although the observed [$\alpha$/Fe] ratios ($\sim 0$)
are lower than the above stars, the ratios can be reproduced with (1) $13-15M_\odot$ supernovae (solid line), (2) $13-40M_\odot$ supernovae with mixing-fallback, or (3) hypernovae (dotted line).
Compared with the above star, a smaller mass cut $M_{\rm cut}$ is favored at a given progenitor mass, in order to obtain larger iron yields and thus lower [$\alpha$/Fe] ratios.
With mixing-fallback, larger $M_{\rm mix}$ and/or larger $f$ values could also make lower [$\alpha$/Fe] ratios.
To be consistent with observed supernovae,
$40M_\odot$ supernova models require unrealistically extended mixing and large fallback ($M_{\rm mix}>8M_\odot$ and $f<0.2$).
Among hypernovae, $25M_\odot$ models give a slightly better fit than $40M_\odot$ models.

{\it SDSSJ161956+170539} --- This star shows a carbon enhancement, and thus the enrichment source should be faint supernovae (dotted line) or faint hypernovae (dashed line).
A large fraction of material in the mixing region falls back onto a remnant, and thus the ejected iron mass is smaller than $M({\rm Fe})=0.001M_\odot$.
If there is no such large fallback, the [C/Fe] ratio becomes too small (solid line).
This is the case for models of other CEMP stars at [Fe/H]$\ltsim -3$, independent of the [$\alpha$/Fe] ratios.
In order to reproduce the observed low [Mg/Fe] ratio, it is necessary that the progenitor mass is lower than that for other CEMP stars, or that there is more extended mixing and larger fallback ($M_{\rm mix} \gtsim 8M_\odot$).
Among $\alpha$ elements, [(Si,Ca)/(O,Mg)] ratios tend to be higher for massive stars.
Therefore, $25-40M_\odot$ supernova/hypernova models may produce [Si/Fe] and [Ca/Fe] ratios larger than the observed ratios.
In any case, the remnant mass is as massive as $M_{\rm rem} \sim 5M_\odot$, which is in the range of black holes.
Although the $M({\rm Fe})$ in Table 1 may be too small to make a star with [Fe/H] $\sim -3$, there are acceptable models that have $M({\rm Fe}) \sim 10^{-4}M_\odot$ with $\chi^{2}/N=0.30$ and $0.86$ for $15$ and $25M_\odot$ models, respectively.

To put a further constraint on the models of enrichment sources with abundance fitting, it is necessary to use elemental abundances of iron-peak elements.
We also apply the abundance fitting for a subset of the sample from \citet{coh13} that have [Fe/H] $<-3$, low [Mg/Fe] $<0.25$, and high signal-to-noise ratios, S/N $>80$.
This set includes four stars (HE2357-0701, HE1416-1032, HE0305-5442, and BS16467-062 with [Mg/Fe] $=0.12, 0.18, 0.22$ and $0.24$, respectively) that do not show carbon enhancement, and one CEMP-low-$\alpha$ star (HE2356-0410 with [Mg/Fe]$=0.11$) that is similar to \citet{caf13}'s CEMP star.
Figure 2 shows the comparison of the elemental abundance patterns between observations and favored models.
NLTE corrections are included for Al observations ($+0.6$, \citealt{coh13}), but not for Na ($\sim -0.2$ to $-0.5$), and we apply a constant shift of $+0.2$ for Cr I \citep{lai08}, but not for Mn I ($\sim +0.3$).

For all of the five low [Mg/Fe] stars, the abundance patterns can be reproduced with core-collapse supernovae, i.e., (1) $13-15M_\odot$ supernovae, (2) $13-40M_\odot$ supernovae with mixing-fallback, or (3) hypernovae, except for the following elements.
Sc, Ti, V, and Co yields depend on the density structure of progenitor stars \citep{tom07}, and can be larger with two-dimensional explosions \citep{tom09}.
Mn and Co yields depend on $Y_{\rm e}$ that can vary due to the neutrino process during the explosion.
Co and Zn yields depend on the explosion energy \citep{ume05,kob06,tom07}.
N, Na, and Al yields depend not only on the existence of rotation, but also the degree of mixing of hydrogen during hydrostatic burning.
In our abundance fitting, Na, Al, K, Sc, and Ti are excluded.

For all five stars, compared with other models, $40M_\odot$ supernova models show stronger odd-Z effects. In particular, [Co/Fe] ratios are much lower than in $13-25M_\odot$ supernova models and hypernova models.
This relative difference in [Co/Fe] among models will remain, although the absolute amounts of Co yields should be increased with multi-dimensional explosions with the neutrino process.
Therefore, $40M_\odot$ supernova models are strongly disfavored from the observed Co abundances.
Among four non-CEMP stars, HE1416-1032 and HE2357-0701 show some nitrogen enhancement with ([N/Fe], [N/C])=(0.37, $>0.80$) and (0.87, 0.61), respectively.
This nitrogen enhancement could be explained with faint supernova/hypernova models.
We should note, however, that the C and N abundances may be affected by internal mixing \cite[e.g.,][]{spi05}.
For the CEMP star, the enrichment source should be faint supernovae/hypernovae.
In most of cases, C/O is lower, Si/O is higher, (Cr,Mn)/Fe are lower, and (Co,Zn)/Fe are higher for $40M_\odot$ hypernova models than for $13-15M_\odot$ supernova models.
This observed high [Zn/Fe] ratio can be realised only with hypernova models, i.e., with even higher explosion energy.
Note that the observed [Mn/Fe] ratio may be higher than in our models, which could be explained with the variation of $Y_{\rm e}$ due to the neutrino process.

There are other low-$\alpha$ EMP stars in the literature. In \citet{yon13}, 13 out of 47 stars show [Mg/Fe] $<0.2$, although Zn abundances are not available.
In \citet{bar05}, 57 out of 253 stars show [Mg/Fe] $<0.2$, although the data quality is not as good as in \citet{coh13} and \citet{yon13}.
Among the 57 stars, 7 stars show high [Zn/Fe] $>0.3$ including HE0547-4539 at [Fe/H] $=-3.01$, indicating hypernovae, while 6 stars do not show such Zn enhancement, and may be consistent with $13-25M_\odot$ supernovae.

\section{Discussion}

The variation of [$\alpha$/Fe] ratios of EMP stars has been known. 
\citet{kob00} commented that some anomalous stars that have low [$\alpha$/Fe] at [Fe/H] $\ltsim-1$ can be explained not by SNe Ia but by $13-15M_\odot$ core-collapse supernovae.
\citet{arg02} showed that the predicted [$\alpha$/Fe] ratios have a too large scatter in stochastic models of chemical evolution with the yields from \citet{nom97}.
In the updated yields of \citet{kob06}, the ejected iron mass $M({\rm Fe})$ was constrained from the observations of supernovae \citep{nom03}.
As a result, $M({\rm Fe})$ was decreased in low-mass supernovae and increased in high-mass supernovae.
Therefore, the intrinsic variation of [$\alpha$/Fe] became much smaller in \citet{kob06} than in \citet{nom97}.
In chemodynamical simulations with the updated yields, the scatter of [$\alpha$/Fe] is comparable to observations (Fig. 13 of \citealt{kob11mw}).

The mixing-fallback effect is naturally expected in the case of hypernovae, which are jet-induced explosions followed by black hole formation \citep{tom07a,tom09}.
In hypernovae, large fallback is possible even for such energetic explosions as $E_{51}\gtsim10$ if the timescale of jet-injection is long \citep{tom07a}.
In supernovae, it is also possible that some degree of mixing occurs through Rayleigh-Taylor instability \citep[e.g.,][]{hac90,jog09}.
\citet{hac91} showed that more massive supernovae tend to undergo less extended Rayleigh-Taylor mixing (i.e., smaller $M_{\rm mix}$) because of the smaller deceleration of the expanding core due to smoother density structures.

The origin of CEMP-low-$\alpha$ is interesting.
If there is a CEMP-low-$\alpha$ star with [Zn/Fe] $\sim$ [Co/Fe] $\sim 0$, that would require a new population, i.e., $\sim 10-20M_\odot$ faint supernovae that form black holes.
For $\ltsim 20M_\odot$ stars, they are believed to form neutron stars, and thus black-hole forming faint supernovae have not been discussed in previous works.

On the other hand, $\sim 10-20M_\odot$ supernovae are not rare.
Because of the smaller ejection mass than that of more massive stars, the contribution from $\sim 10-20M_\odot$ supernovae can appear at various metallicities in the system where chemical enrichment took place inhomogeneously.
One example may be dwarf spheroidal galaxies (dSphs).
It is known that stars in dSphs show lower [$\alpha$/Fe] than the stars in the solar neighborhood at a given metallicity \citep[e.g.,][]{tol09},
and some EMP stars in dSphs also show low [$\alpha$/Fe] \citep{aok09}.
This is sometimes explained by the contribution from SNe Ia with a longer timescale of star formation (if the SN Ia lifetimes do not depend on the progenitor metallicity).
However, the elemental abundance patterns, in particular the [Mn/Fe] ratios as low as halo stars \citep{nor12}, are not consistent with the enrichment from SNe Ia.
In small systems such as dSphs, it is likely that the star formation efficiency is low and thus the sampling of massive stars is incomplete, which results in the missing contribution from massive stars at $\gtsim20M_\odot$ (\citealt{kob06,nom13}, see also \citealt{mcw13}).
In the Milky Way Galaxy, similar abundance patterns was reported by \citet{nis10,nis11}, which tend to be accreted halo stars from their kinematics.
\citet{ish12} also showed that stars with the Galactic outer halo kinematics tend to have low [$\alpha$/Fe] ratios.
For such an accretion component, incomplete sampling of massive stars is possible, although the metallicity is higher than that of EMP stars.
The abundance patterns of these stars are also consistent with our $13-25M_\odot$ supernova models.

Abundance fitting is a new approach to study the properties of supernovae.
Statistical studies with a large homogeneous sample will be able to answer the following questions:
What is the major source of chemical enrichment in the early Universe?
Namely, what were the distribution functions of progenitor mass, ejected iron mass, and explosion energy of supernovae?
Are the mass-energy and mass-iron mass relations of nearby supernovae reproduced?
What is the mass function of the remnants of the first stars?

\section{Conclusions}

In the early stages of chemical enrichment, the interstellar medium is supposed to be highly inhomogeneous, so that the properties of the first supernovae can be directly extracted from the comparison between the observed elemental abundances and nucleosynthesis yields.
We show that the low [$\alpha$/Fe] ratios recently found for a small fraction of extremely metal-poor stars can be naturally explained with the nucleosynthesis yields of core-collapse supernovae, i.e., (1) $13-25M_\odot$ supernovae, or (2) hypernovae.
If we allow an enhanced mixing and a large fallback, $40M_\odot$ supernova models (3) could be consistent with the observed low [$\alpha$/Fe] ratios.
For the case without carbon enhancement, the ejected iron masses of these favored models (1-3) are normal, $M({\rm Fe})=0.05-0.15M_\odot$ for supernovae and $M({\rm Fe})=0.1-1.4M_\odot$ for hypernovae, consistent with observed light curves and spectra of nearby supernovae.
The first source (1) has been included in the standard set of nucleosynthesis yields that have been applied to galactic chemical evolution models, while the other sources are different from those in galactic chemical evolution models and rarer.

On the other hand, the carbon enhancement requires much smaller iron production, and the low [$\alpha$/Fe] of carbon enhanced metal-poor stars can also be reproduced with faint supernovae or faint hypernovae.
The ejected iron mass is $M({\rm Fe})<0.001M_\odot$, much smaller than normal supernovae.
These enrichment sources are similar to those proposed for typical carbon-enhanced EMP stars and DLAs with [$\alpha$/Fe]$\gtsim0.5$, but the progenitor mass is as low as $13-25 M_\odot$, or more extended mixing and larger fallback occur in $25-40M_\odot$ stars.
The former case implies that $\ltsim 25M_\odot$ stars may form black holes.

Iron-peak element abundances, in particular Zn abundances, are important to put further constraints on the enrichment sources.
$25-40M_\odot$ supernova (not hypernova) models may disagree with the observed high Co abundances of low [$\alpha$/Fe] stars.
The frequency of these sources should be examined with future galactic archaeology surveys.

\acknowledgments
We would like to thank E. Caffau and P. Bonifacio for providing their results prior to publication.
We are grateful to A. Bunker for fruitful discussion.
This work has been supported in part by WPI Initiative, MEXT, Japan, and by the Grant-in-Aid for Scientific Research of the JSPS (18104003, 20540226) and MEXT (19047004, 22012003).

\begin{figure}
\center 
\includegraphics[width=16cm]{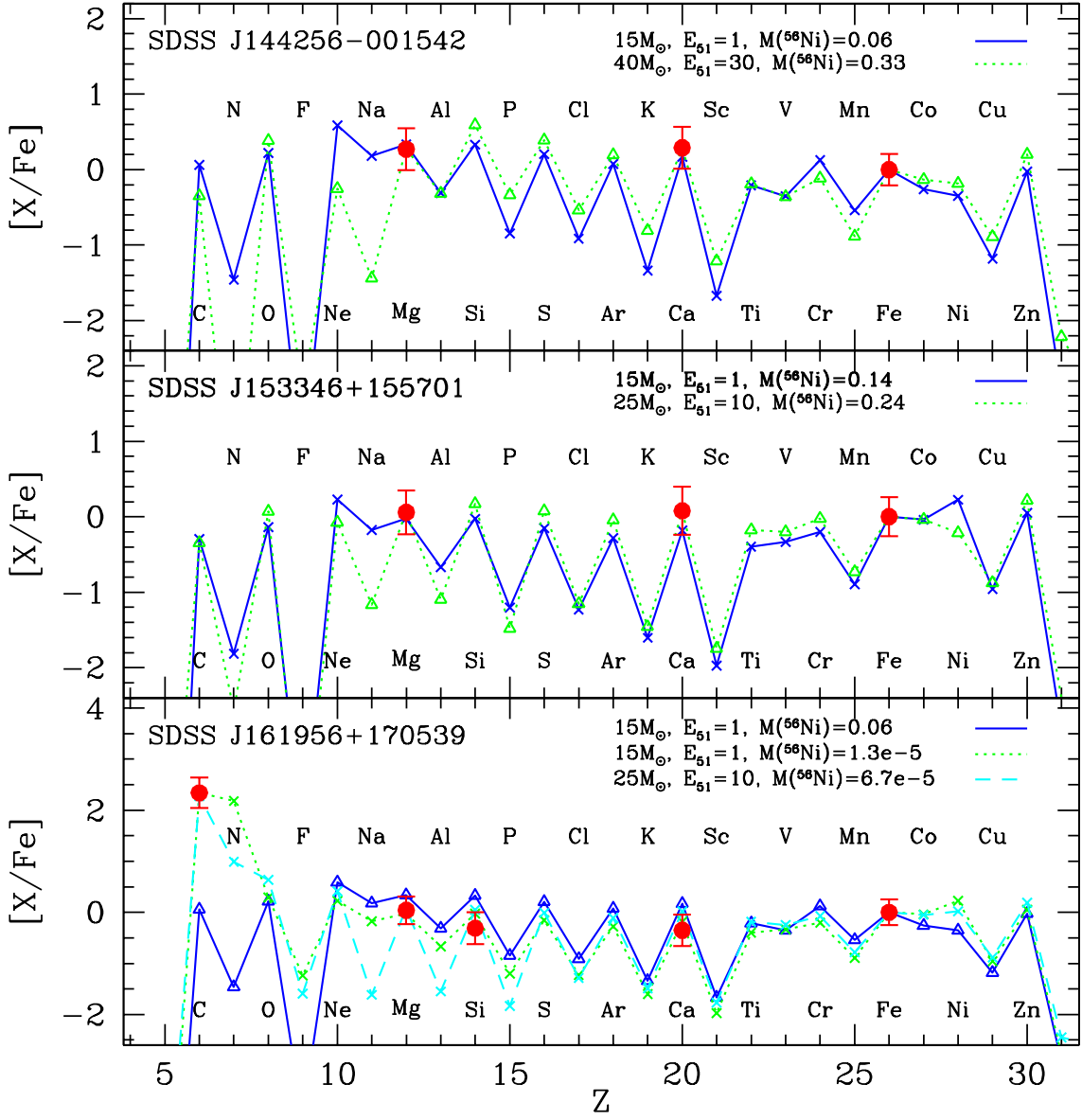}
\caption{
Elemental abundance patterns of three EMP stars with (top and middle panels) and without (bottom panel) carbon enhancement from \citep{caf13}.
In the top and middle panels, the solid and dotted lines show the nucleosynthesis yields of a low-mass supernova without mixing-fallback and hypernova, respectively.
In the bottom panel, the solid, dotted, and dashed lines are for a normal supernova without mixing-fallback, faint supernova, and faint hypernova, respectively.
}
\end{figure}

\begin{figure}
\center 
\includegraphics[width=16cm]{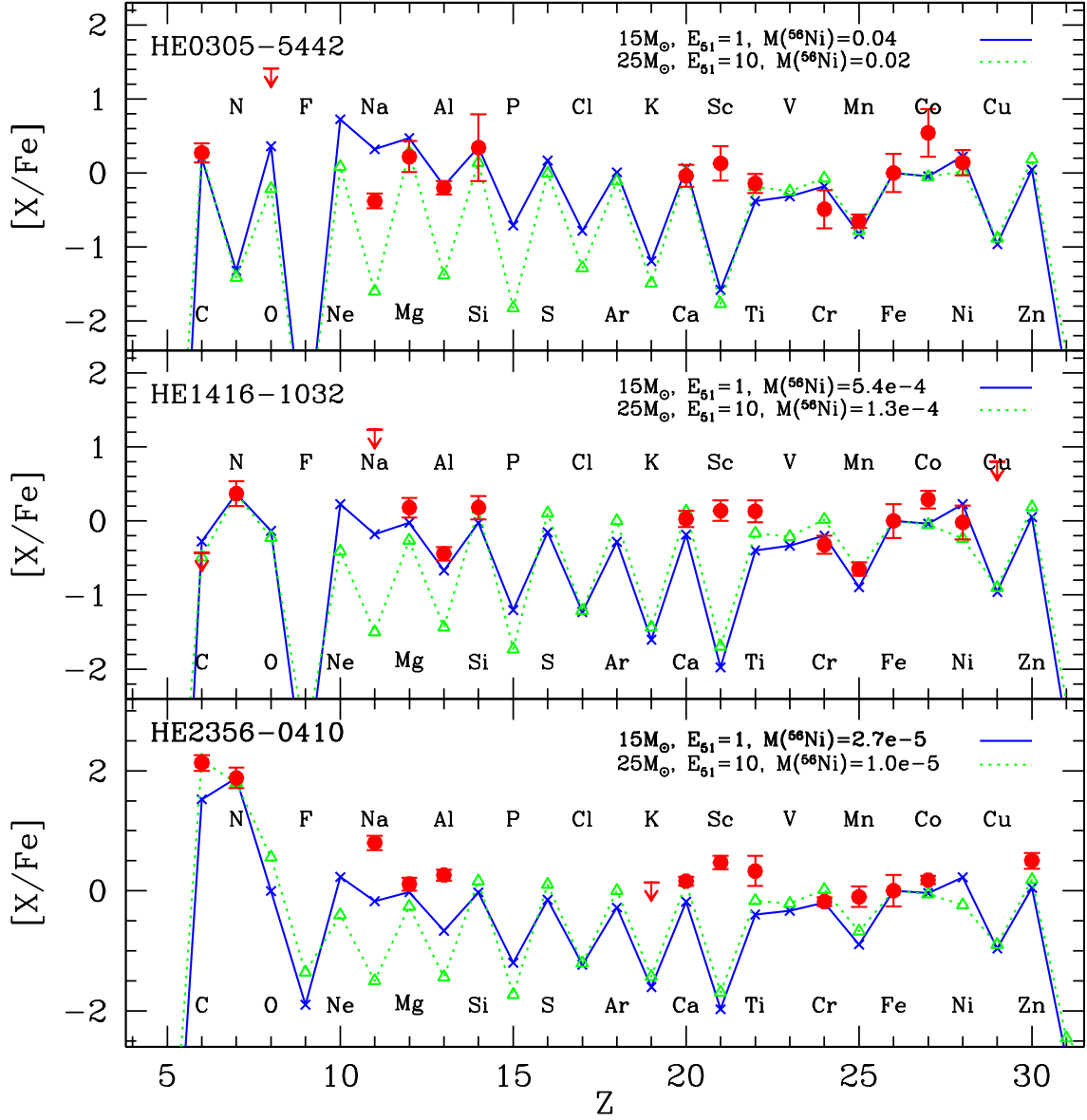}
\caption{
Elemental abundance patterns of three EMP stars with (top and middle panels) and without (bottom panel) carbon enhancement from \citet{coh13}.
In the top and middle panels, the solid and dotted lines show the nucleosynthesis yields of a low-mass supernova and hypernova, respectively.
In the lower panel, the solid and dotted lines are for a faint supernova and faint hypernova, respectively.
All models are with mixing-fallback.
}
\end{figure}

\end{document}